\documentclass[twocolumn,prb,superscriptaddress]{revtex4}

\usepackage{graphicx}
\usepackage{dcolumn}
\usepackage{bm}
\usepackage{amssymb}

\def\ns{$\times$10$^{10}$ cm$^{-2}$}
\def\mo{$\times$10$^6$ cm$^2$/Vs }
\def\sigxx{Re[$\sigma_{xx}(f)$]}
\def\fpk{$f_\mathrm{pk}$}

\begin{document}

\preprint{}

\title{Melting of a 2D Quantum Electron Solid in High Magnetic Field}

\author{Yong P. Chen}
\altaffiliation[Current address: ]{Richard Smalley Institute for Nanoscale Science and Technology and Dept. of Physics, Rice University, Houston TX 77005, USA}
\affiliation{Department of Electrical Engineering, Princeton University,
Princeton, NJ 08544}
\affiliation{National High Magnetic Field Laboratory, 1800 E.~Paul Dirac Drive, 
Tallahassee, FL 32310}
\author{G. ~Sambandamurthy}
\affiliation{National High Magnetic Field Laboratory, 1800 E.~Paul Dirac Drive, 
Tallahassee, FL 32310}
\affiliation{Department of Electrical Engineering, Princeton University,
Princeton, NJ 08544}
\author{Z.~H.~Wang}
\affiliation{Department of Physics, Princeton University,
Princeton, NJ 08544}
\affiliation{National High Magnetic Field Laboratory, 1800 E.~Paul Dirac Drive, 
Tallahassee, FL 32310}
\author{R. ~M.~ Lewis}
\altaffiliation[Current address: ]{Dept. of Physics, University of Maryland, College Park MD 20742, USA}
\affiliation{National High Magnetic Field Laboratory, 1800 E.~Paul Dirac Drive, 
Tallahassee, FL 32310}
\affiliation{Department of Electrical Engineering, Princeton University,
Princeton, NJ 08544}
\author{L.\ W.\
Engel}
\affiliation{National High Magnetic Field Laboratory, 1800 E.~Paul Dirac Drive, 
Tallahassee, FL 32310}
\author{D.\ C.\ Tsui}
\affiliation{Department of Electrical Engineering, Princeton University,
Princeton, NJ 08544}
\author{P.\ D.\ Ye}
\altaffiliation[Current address: ]{School of Electrical and Computer Engineering, Purdue University, West Lafayette, IN 47907, USA}
\affiliation{National High Magnetic Field Laboratory, 1800 E.~Paul Dirac Drive, 
Tallahassee, FL 32310}
\affiliation{Department of Electrical Engineering, Princeton University,
Princeton, NJ 08544}
\author{L.\ N.\ Pfeiffer}
\affiliation{Bell Laboratories, Lucent Technology, Murray Hill, NJ 07974}
\author{K.\ W.\ West}
\affiliation{Bell Laboratories, Lucent Technology, Murray Hill, NJ 07974}


\maketitle


{\bf 
The melting temperature ($T_m$) of a solid is generally determined by the pressure applied to it, 
or indirectly by its density ($n$) through the equation of state.  This remains true even for helium solids\cite{wilk:67}, where quantum effects often lead to unusual properties\cite{ekim:04}.  In this letter we present experimental evidence to show that for a two dimensional (2D) solid formed by electrons in a semiconductor sample under a strong perpendicular magnetic field\cite{shay:97} ($B$), the $T_m$ is not controlled by $n$, but effectively by the \textit{quantum correlation} between the electrons through the Landau level filling factor $\nu$=$nh/eB$.  Such melting behavior, different from that of all other known solids (including a classical 2D electron solid at zero magnetic field\cite{grim:79}), attests to the quantum nature of the magnetic field induced electron solid.  Moreover,  we found the $T_m$ to increase with the strength of the sample-dependent disorder that pins the electron solid.  }

Electrons are expected to crystalize into a solid (so called ``Wigner crystal"\cite{wign:34} ) when the 
(Coulomb) interaction energy between the electrons sufficiently dominates over the kinetic energy. 
One example of such an electron solid was found in a very low density ($n$) two dimensional electron 
system (2DES) realized on helium surfaces\cite{grim:79} (at zero magnetic field).  Because of the low $n$, the zero-point motion (given by the Fermi energy $E_f=nh^2/2\pi m$ where $m$ is the electron mass) is negligibly small and at finite temperatures ($T$), as in the experiment\cite{grim:79}, the kinetic energy originates mainly from the \textit{classical} thermal motion ($k_BT$).   
The melting of such a ``classical" 2D electron solid is
determined only by the competition between the thermal kinetic energy and Coulomb interaction 
($e^2\sqrt{n}/4\pi\epsilon$) [MKS unit is used exclusively in this paper] and is thought to be describable by the Kosterlitz-Thouless theory 
of 2D melting\cite{thou:78}.  Experimentally, the melting was found to occur\cite{grim:79} at $T_{cm}=e^2\sqrt{n}/(4\pi\epsilon k_B\Gamma)$ (where $\epsilon$ is the dielectric constant and $k_B$ the Boltzman constant) with $\Gamma$$\sim$130, in excellent agreement with theoretical calculations\cite{morf:79,chui:91}. 

The 2DES as realized in high-quality GaAs/AlGaAs structures in our experiment has relatively high $n$, thus (in the absence of magnetic fields) the zero-point motion ($E_f$) is significant and the 2DES does not solidify even at $T$=0. However, it is well known that a sufficiently strong perpendicular magnetic field ($B$) tends to suppress the kinetic energy of 2D electrons and induce the solidification\cite{lozo:75,fuku:79}. 
 On the other hand, at finite $B$ the motion of electrons is quantized into
Landau levels (LL) and delicate many-body quantum correlations\cite{laug:83,kive:87}
among electrons can cause the 
2DES to condense into fractional quantum Hall (FQH)\cite{tsui:82} liquid states at certain rational fractional values of Landau level filling factor $\nu=nh/eB$. Experimentally,
the ``magnetic field induced electron solid" (MIES)\cite{shay:97} forms at sufficiently small $\nu$, following the termination of FQH states at 
low $T$. It is deformed and pinned by disorder in the system, thus an insulator (in DC transport), and has a characteristic resonance (see, for example, Ref.~14\nocite{pdye:02}) in its frequency ($f$) dependent, real diagonal conductivity (\sigxx) due to the ``pinning mode" of domains of the elastic solid oscillating collectively around the disorder\cite{chit:9802, fert:99,fogl:00}. 

The melting of such a MIES has been studied in various 
experiments\cite{glat:90,vgol:90, fwil:91, paal:92,egol:92,kuku:9394} and
it was commonly presumed\cite{glat:90, fwil:91,egol:92,chui:91} that at a fixed $\nu$ the melting 
should be similar to that of a classical electron 
solid (the expected exact ground state for a 2DES at infinite $B$) 
and thus $T_m$ would be determined by $n$, as 
$T_{cm}$ is.    In this letter, we show unambiguously that this is not the case, that is, the melting of a
MIES does \textit{not} behave as a classical electron solid, and in fact in any given sample (whose intrinsic disorder is fixed), $T_m$ is determined by $\nu$, not $n$, and is unrelated to $T_{cm}$.  

In our experiments we have studied the $T$-dependence of the pinning mode resonance of the 
MIES in its \sigxx\ spectrum measured by microwave spectroscopy\cite{enge:93}.  
No resonance is observed when $T$ is raised above some characteristic 
$T_m$, taken as the melting $T$ of the electron solid. By systematically measuring $T_m$ while varying 
\textit{both} $n$ and $B$, we found that within the experimental resolution, $T_m$ in a given sample is
only a function of $\nu$ (i.e., $T_m(n,B)=T_m(n/B)$) down to $\nu$ as small as $\sim$0.05 attained in our experiments.  Although $T_m$ is generally sensitive to $n$ at fixed $B$,  $T_m$ is \textit{insensitive} to $n$ at fixed $\nu$.  Since $\nu=nh/eB=2(l_B/r)^2$, where the magnetic length $l_B=\sqrt{\hbar/eB}$ is a measure of the size of the single electron wavefunction and $r=1/\sqrt{\pi n}$ is the mean separation between the electrons, our findings reflect the quantum nature of the 2D electron solid formed at finite $B$ and demonstrate that its melting ($T_m$) is effectively controlled by the inter-electron quantum correlation, which depends on $l_B/r$.

We have studied two different 2DES samples.  Sample 1 is a GaAs/AlGaAs heterojunction. Sample
2 is a 15nm-wide AlGaAs/GaAs/AlGaAs quantum well (QW).  By backgating and/or different cooldowns, the electron densities ($n$) in both samples can be tuned to various extents (to be specified below).   Typically at their respective as-cooled $n$, sample 1 has mobility $\mu$$\sim$6\mo , sample 2 has $\mu$$\sim$1\mo.

\begin{figure}
\includegraphics[width=8.2cm]{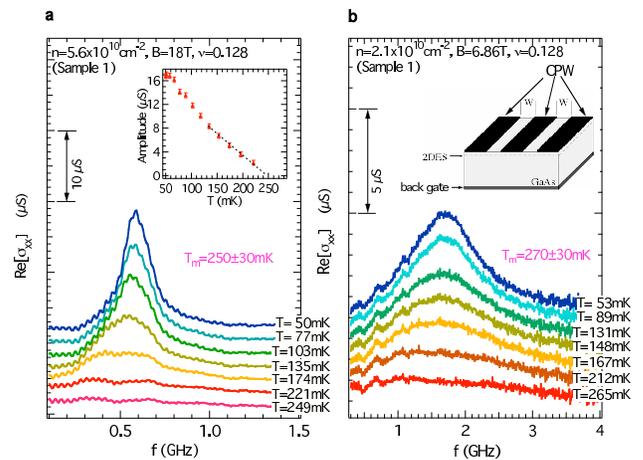}
\caption{\label{melt_fig1} Temperature ($T$) dependence of the microwave spectra of the 2D electron solid measured from sample 1 at two different densities ($n$) with the same Landau filling factor ($\nu$=$nh/eB$). \textbf{a}, $T$-dependence of the spectra at $n$=5.6\ns\  with $\nu$=0.128 (magnetic field $B$=18 Tesla).  Spectra at a series of representative $T$'s are shown and offset for clarity. The pinning resonance of the electron solid observed at $T$=50 mK is seen to weaken with increasing $T$ and disappear at $\sim$250 mK, taken as the melting temperature ($T_m$) of the electron solid. Inset shows the amplitude of the resonance extrapolates to zero at the similar $T_m$.  \textbf{b}, $T$-dependence of the spectra at $n$=2.1\ns\ with $\nu$=0.128 ($B$=6.86 Tesla). The low $T$ resonance disappears at a $T_m$ similar to that shown in \textbf{a} (n=5.6\ns), despite here that $n$ has been reduced by more than a factor of 2 (using a negative voltage between a backgate and the 2DES).   Inset shows a schematic of the sample. The dark regions on the top surface represent the coplanar wave guide (CPW). The backgate allows the \textit{in-situ} change of $n$.}
\end{figure}

The inset of Fig.~\ref{melt_fig1}(b) shows a schematic local cross section
of a typical sample (not to scale). A metal film coplanar waveguide (CPW), lithographically deposited on the  
surface, enables the measurement of \sigxx \ of the 2DES. A 
network analyzer generates a microwave signal propagating along the CPW and coupling capacitively to the 
2DES (located some 0.1-0.5 $\mu$m below the surface).  The relative power absorption ($P$) by the 2DES is measured and can be related\cite{enge:93} to Re[$\sigma_{xx}$] of the 2DES as $P=\exp((2lZ_0/w)\mathrm{Re}[\sigma_{xx}])$, where $l$ and $w$ are the total length and slot width of the CPW respectively and
$Z_0$=50$\Omega$ is the CPW characteristic impedance.  CPW of meander shapes\cite{enge:93}
are commonly used to obtain larger geometric ratios ($2l/w$), and therefore to 
increase the strength of the signal ($P$).
The sample is mounted on a metal block that is kept in good thermal equilibrium with the mixing 
chamber of a dilution refrigerator during the measurements. The microwave input is kept in the low power limit, by reducing the power till the signal ($P$) no longer changes.  A negative voltage between a backgate (located $\sim$200 $\mu$m below the surface) and the 2DES 
enables \textit{in-situ} reduction of $n$ from the as-cooled values. 

Fig.~\ref{melt_fig1} shows $T$-dependence of the microwave resonance of the electron solid in 
sample~1 and the determination of $T_m$ at two different values of $n$ with $\nu$ fixed at a representative value of 0.128.  In Fig.~\ref{melt_fig1}a, $n$=5.6\ns\ ($B$=18 T) and the \sigxx \ spectrum displays a clear resonance with peak frequency (\fpk) near 600 MHz at low $T$($\sim$50 mK).   As $T$ is increased, the resonance weakens.  Its \fpk\  at elevated $T$ also decreases slightly from the base $T$ (50 mK) value, but by no more than 20\%, indicating that there is no significant loss of pinning of the electron solid due to the increase of the temperature.  The resonance disappears into noise background at $\sim$250 mK, which we take as the melting $T$ ($T_m$) of the electron solid.  The inset of Fig.~\ref{melt_fig1}a shows that the resonance amplitude (obtained from a Lorentzian fit) extrapolates to zero at a similar $T_m$.  
In Fig.~\ref{melt_fig1}b, $n$ has been reduced to 2.1\ns\ (using a backgate
voltage of $-$300 V) while $B$ is also reduced to 6.86 T to keep $\nu$ the same value as in Fig.~\ref{melt_fig1}a.  The lower $n$ reduces the electron-electron interaction relative to the electron-disorder interaction, and the pinning resonance of the electron solid now occurs (at low $T$) near 1.6 GHz, which is significantly higher than that in Fig.~\ref{melt_fig1}a.  However, the resonance disappears above a $T_m$$\sim$ 270 mK, which is similar to the $T_m$ in Fig.~\ref{melt_fig1}a, within the experimental uncertainty in determining $T_m$ (typically $\sim$10\%).

\begin{figure}
\includegraphics[width=8.2cm]{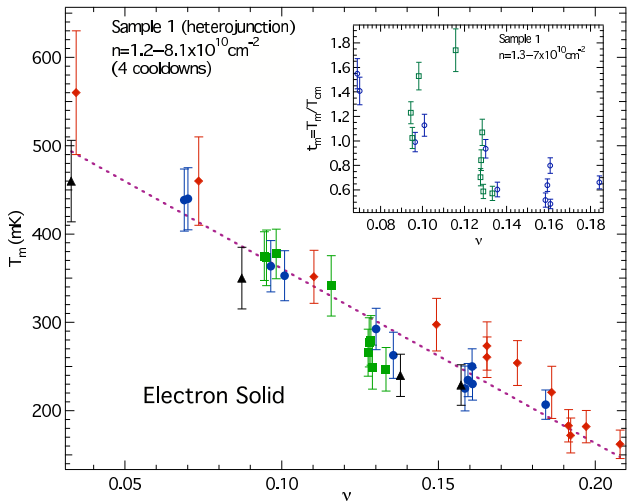}
\caption{\label{melt_pd} ($T_m$,$\nu$) phase diagram for the electron solid in sample 1. The $T_m$'s are measured in
a total of 4 cooldowns (shown as diamonds, circles, squares and triangles), over a wide range of densities ($n$=1.2$-$8.1\ns) and magnetic fields. Within the experimental 
uncertainty,  $T_m$ versus $\nu$ gives rise to a well-defined melting curve of the electron solid.  The dashed line is a guide to the eye, obtained by a linear fit through all the data. Typical error bars in $T_m$ are less than 10\%. The inset shows the ``reduced" $t_m$ versus $\nu$ from two cooldowns.  $t_m$ is defined as $T_m$ normalized by the classical 2D electron solid melting temperature $T_{cm}=e^2\sqrt{n}/(4\pi\epsilon k_B\Gamma)$ (where we take $\Gamma$=127).  $t_m$ versus $\nu$ does not result 
in a well-defined melting curve, indicating that the meting temperature $T_m$ is not determined by $n$ or $T_{cm}$.}
\end{figure}

We have measured $T_m$ in 4 different cooldowns of sample 1 for many combinations of $n$ and $B$,
at which a resonance from the MIES can be detected. We plot all these $T_m$ data as a function of $\nu$ in Fig.~\ref{melt_pd}. Sample 1 has a particularly large range of tunable $n$, from $\sim$1.2$-$8.1\ns, covering a $\nu$ range from $\sim$0.21 down to $\sim$0.03.  Generally, we found $T_m$ to be sensitive to $n$ or $B$ separately (when fixing one and changing the other). On the other hand,  near similar $\nu$ (changing both $n$ and $B$ while fixing $n/B$), we have always found $T_m$ to be insensitive to $n$ (or $B$), within the experimental errors in $T_m$.  Different cooldowns can vary $T_m$ by up to $\sim$15\% (at similar $\nu$) but this does not affect our conclusion. We thus find $T_m$ to be mainly determined by $\nu$, so that $T_m$ vs $\nu$ as plotted in Fig.~\ref{melt_pd} defines a melting curve for the electron solid in sample 1.   A linear fit of $T_m$ vs $\nu$ gives a guide to the eye shown as the dashed line, which lies within 20\% from all the ($T_m$, $\nu$) data points and within 10\% from a majority (70\%) of them. In the inset of Fig.~\ref{melt_pd} we plot the ``reduced"\cite{chui:91,glat:90} $t_m=T_m/T_{cm}$ from two similar cooldowns versus $\nu$,  where $T_{cm}=e^2\sqrt{n}/(4\pi\epsilon k_B\Gamma$) is the melting $T$ of a classical 2D electron solid defined earlier (with the value\cite{morf:79,chui:91} $\Gamma$=127).  In contrast to $T_m$, $t_m$  can vary significantly (sometimes by a factor of 3) at similar $\nu$.
Thus $t_m$ versus $\nu$ does not give a well defined melting curve for the electron solid, confirming that $T_m$ is not
determined by $n$ or $T_{cm}$.  We have also checked that neither $T_m$ nor $t_m$ plotted against either $n$ or $B$ gives any 
well defined melting curve.   We further notice that our measured $T_m$ is also unrelated to the hypothetical ``thermal depinning"  temperature  ($T_\mathrm{depin}$) of domains\cite{fert:99,fogl:00,chit:9802} of the electron crystal. In fact, one can estimate $T_\mathrm{depin}$$\sim$$(1/k_B)m_e\omega^2_0\xi^2(L/a)^2$$\propto$$n^{3/2}$ 
(where $\omega^2_0$=$2\pi f_\mathrm{pk}eB/m_e$$\propto$
$n^{-3/2}$ characterizes the ``effective pinning potential"\cite{fogl:00}(we note that $\omega_0$ is also much larger than $k_BT_m/\hbar$), $\xi$ is the disorder correlation length,  $L$$\propto$$n$ is the Larkin domain size\cite{fert:99,fogl:00,chit:9802} and $a$ is the electron crystal lattice constant), in contrast to $T_m$, which is determined by $\nu$, and as seen in Fig.~~\ref{melt_pd},
typically decreases with increasing $n$ if $B$ is fixed.  

\begin{figure}
\includegraphics[width=8.2cm]{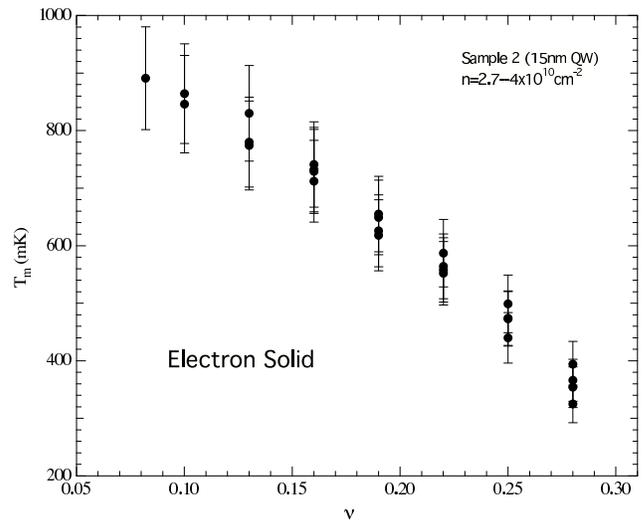}
\caption{\label{melt_nqw} ($T_m$,$\nu$) melting curve of the electron solid in sample 2,
a narrow QW of width 15 nm. Sample 2 has a tunable $n$=2.7-4\ns\ and enters an electron solid phase at $\nu$$<$ 0.3. }
\end{figure}

Fig.~\ref{melt_nqw} shows the ($T_m$,$\nu$) melting curve measured on sample 2 (15nm-wide 
QW).  Sample 2 has a tunable $n$=2.7-4\ns. Likely due to the relatively narrow confinement of the 2DES 
in the QW, sample 2 enters the solid phase for $\nu$$<$0.3 and the typical \fpk\ observed is $\sim$6-8 GHz. In this sample we found $T_m$ again to be controlled by $\nu$, though the value of $T_m$ is higher than in sample 1 at similar $\nu$.   

In each of the two samples, at a fixed high-$B$, $T_m$ typically increases with decreasing $n$. This is opposite to the classical behavior, where $T_{cm}$ decreases with decreasing $n$. At a fixed $\nu$, our observed $T_m$ is insensitive to $n$, and $T_m$ in each sample is a well defined function of $\nu$. Recent theories\cite{hyi:98,chan:05} have suggested that many-body quantum correlation between electrons can still be important in the solid phase terminating the FQH states
at high $B$.  The parameter that captures such quantum correlation is the Laudau filling factor $\nu$=$2(l_B/r)^2$, which also determines the (single-particle) wavefunction overlap\cite{maki:83} between neighboring electrons.  This overlap ($I_s$), given\cite{maki:83} by $e^{-(2/\sqrt{3})\pi/\nu}$, is in fact quite small in the high-$B$, low-$\nu$ regime of the MIES.   Our data show that $T_m$ is quite strongly dependent on $\nu$ (for example, $T_m$ changes by almost a factor of 2 when $\nu$ is changed by a factor of 2). Therefore the dependence of  $T_m$ on $\nu$ is unlikely to result from a $T_m$ that is determined by the Coulomb Hartree energy $E_H$ (which is shown\cite{maki:83} to have very weak $\nu$-dependence in this low-$\nu$ regime, where the wavefunction overlap $I_s$ has negligible contribution to $E_H$).  Our findings indicate that the melting of MIES is effectively controlled by the inter-electron quantum correlation\cite{maki:83}, through $\nu$.   The well-defined ($T_m$,$\nu$) melting curve we obtained constitutes the phase boundary between a \textit{quantum solid} and a correlated quantum liquid\cite{chit:9802} in each sample.   

The $T_m$ we measured in both samples are of similar \textit{order of magnitude} to those in other samples measured previously with various experimental techniques\cite{glat:90,vgol:90, fwil:91,paal:92,egol:92,kuku:9394}.  We have noticed that at similar $\nu$, sample 2 (narrow QW) has higher $T_m$ than sample 1 (heterojunction).  It has been suggested\cite{fert:99} that the relevant disorder in semiconductor samples that pins the MIES comes mostly from the interfaces vertically confining the 2DES (this is consistent with the relatively high \fpk\  observed in the narrow QW sample).  Our findings thus suggest that a 2D electron solid subject to stronger pinning disorder melts at higher $T_m$, an effect that has been predicted earlier\cite{tsuk:77}.  We notice, interestingly, that studies on the effects of pinning disorder or geometric confinement\cite{chri:01} on $T_m$ in other solids usually find the opposite behavior.  For example, both the vortex solid in a high-temperature superconductor subjected to artificial pinning centers\cite{paul:00} and a helium solid in a porous glass\cite{beam:83} show a depression of $T_m$ with the added disorder.  Disorder is generally unavoidable in semiconductor samples, and as seen here can strongly influence the \textit{values} of $T_m$ measured, which may not be those of an ideal, clean Wigner crystal.    On the other hand,
the observation that in any given sample (thus with fixed intrinsic disorder),  $T_m$ is not sensitive to $n$ (nor to $B$) if $\nu$=$nh/eB$ is fixed (but is sensitive to $n$ if $B$ is fixed) is difficult to explain in a classical picture of, for example, interplay of disorder and $n$ (screening). The fact that $\nu$ being the controlling variable for $T_m$ is found in samples of quite different disorder (including sample 1, in which one can be infer from the observed pinning resonance that the Wigner crystal there can possess substantial correlation length\cite{pdye:02,fert:99,fogl:00,chit:9802}) suggests that such a behavior ($T_m$
being controlled by $\nu$) arises from some mechanism in which many-body quantum correlation between electrons may be important. The exact role of disorder (perhaps involving its interplay with quantum correlation) in determining the value and behavior of $T_m$ remains to be better understood.  


It is also interesting to compare the quantum nature of our 2D electron solid to the quantum 
solids of helium. In a helium solid, the size of atoms is fixed by nature and the quantum parameter is 
the De Boer parameter\cite{wilk:67} $\Lambda$$\sim$$h/(a\sqrt{Mv})$ (where $M$ is the atomic mass,
 $a$ the inter-atomic distance and $v$ the inter-atomic potential strength), which is fixed 
 at fixed $n$.    The $T_m$ of a helium solid is only determined by $n$.   In the case of a 2D
 electron solid formed in high $B$, the size of single electron wavefunction ($l_B$) is readily
 tunable by $B$, independently of $n$. The quantum parameter here is $\nu$=$nh/eB$ and we have found that $T_m$ of such an electron solid in a given sample (with fixed disorder) is controlled by $\nu$ rather than $n$.  

\begin{description}
\item[Acknowledgement] Financial support of this work was provided by  DOE grant no. DE-FG02-05ER46212,  the NHMFL in-house research program, and AFOSR.  The spectroscopy measurements were performed at the National High Magnetic Field Laboratory, which is supported by NSF Cooperative Agreement No. DMR-0084173 and by the State of Florida.  We thank Glover Jones, Tim Murphy and Eric Palm at NHMFL for experimental assistance.  We also thank S. T. Chui, R. L. Willett,  Kun Yang and C. C. Yu for inspirational discussions, and thank H. A. Fertig and P. B. Littlewood in particular for discussions regarding the thermal depinning temperature of the Wigner crystal. 

\item[Competing Interests] The authors declare that they have no
competing financial interests.

\item[Correspondence] Correspondence should be addressed to Y.P.C.~(email: yongchen@rice.edu).
 
\end{description}

\end{document}